\documentclass[12pt, letterpaper]{article}
\usepackage[top=1in, bottom=1.5in, left=1in, right=1in]{geometry}

\usepackage[utf8]{inputenc}
\usepackage{aas_macros}
\usepackage{booktabs}
\usepackage{hyperref}
\hypersetup{
    colorlinks=true,
    linkcolor=blue,
    citecolor=blue,
    allcolors=blue
}
\usepackage{color}
\usepackage{graphicx}
\usepackage{caption}

\newcommand{\blue}{\color{black}}
\newcommand{\red}{\color{black}}
\newcommand{\green}{\color{black}}

\title{Optimizing the LSST Observing Strategy for Dark Energy Science: DESC Recommendations for the Deep Drilling Fields and other Special Programs}

\author{
\parbox{0.9\textwidth}{
\normalsize
Daniel~M.~Scolnic$^{1}$,
Michelle~Lochner$^{2,3}$,
Phillipe~Gris$^{4}$,
Nicolas Regnault$^{5}$,
Ren\'ee~Hlo\v{z}ek$^{6,7}$,
Greg~Aldering$^{8}$,
Tarek~Allam Jr$^{9}$,
Humna Awan$^{10}$,
Rahul~Biswas$^{11}$,
Jonathan~Blazek$^{12,13}$,
Chihway Chang$^{14}$,
Eric~Gawiser$^{10}$,
Ariel Goobar$^{15}$,
Isobel~M.~Hook$^{16}$,
Saurabh~W.~Jha$^{10}$,
Jason~D.~McEwen$^{9}$,
Rachel~Mandelbaum$^{17}$,
Phil~Marshall$^{18}$,
Eric~Neilsen$^{19}$,
Jason~Rhodes$^{20}$,
Daniel Rothchild$^{21}$,
Ignacio~Sevilla-Noarbe$^{22}$,
An\v{z}e~Slosar$^{23}$,
Peter~Yoachim$^{24}$
\begin{center} (The LSST Dark Energy Collaboration) \end{center}
}}

\date{}

\begin{document}

\maketitle
\begin{abstract}

{\blue We review the measurements of dark energy enabled by observations of the Deep Drilling Fields (DDFs) and the optimization of survey design for cosmological measurements. This white paper is the result of efforts by the LSST DESC Observing Strategy Task Force (OSTF), which represents the entire collaboration, and aims to make recommendations on observing strategy for the DDFs that will benefit all cosmological analyses  with  LSST.  It  is  accompanied  by  the  DESC-WFD  white  paper  (Lochner et al.). {\green We argue for altering the nominal deep drilling plan to have $>6$ month seasons, interweaving $gri$ and $zy$ observations every 3 days with 2, 4, 8, 25, 4 visits in $grizy$, respectively.  These recommendations are guided by metrics optimizing constraints on dark energy and mitigation of systematic uncertainties, including specific requirements on total number of visits after Y1 and Y10 for photometric redshifts (photo-$z$) and weak lensing systematics. We specify the precise locations for the previously-chosen LSST deep fields (ELAIS-S1, XMM-LSS, CDF-S, and COSMOS) and recommend Akari Deep Field South as the planned fifth deep field in order to synergize with Euclid and WFIRST.}  Our recommended DDF strategy uses $6.2\%$ of the LSST survey time.  We briefly discuss synergy with white papers from other collaborations, as well as additional mini-surveys and Target-of-Opportunity programs that lead to better measurements of dark energy.

}
\end{abstract}
\newpage
\section{White Paper Information\label{sec:WPinfo}}
Please provide contact information (name and email address) of the appropriate author(s) for this white paper.\\
{\blue
Dan Scolnic, \url{dscolnic@kicp.uchicago.edu}\\
Michelle Lochner, \url{dr.michelle.lochner@gmail.com}\\
The LSST Dark Energy Science Collaboration\\}

\begin{enumerate} 
\item {\bf Science Category:} 
{\blue
\begin{itemize}
\item{The Nature of Dark Matter and Understanding Dark Energy} 
\item{Exploring the Changing Sky} 
\end{itemize}
}
\item {\bf Survey Type Category:} 
   \begin{itemize}
   \item This paper focuses on the Deep Drilling Field special program, but remarks on other Mini-surveys and Target-of-Opportunity observations.
    \end{itemize}

\item {\bf Observing Strategy Category:} 
    \begin{itemize} 
    
  {\red \item An integrated program with science that hinges on the combination of pointing and detailed 
	observing strategy. }
   
    \end{itemize}  
\end{enumerate}

\clearpage
\section{Scientific Motivation\label{sec:motivation}}

\begin{footnotesize}
\end{footnotesize}
{\blue

\indent The Deep Drilling Field (DDF) mini-survey is critical for enabling the core science of the LSST Dark Energy Science Collaboration (DESC).  For DESC, the {\green DDF} mini-survey has three drivers: cosmology with Type Ia Supernovae (SNIa), synergy with other surveys, and calibration of {\green primary} systematic uncertainties for multiple cosmological probes.  As shown in \autoref{fig:DEFoM}, the DDFs are needed for DESC to reach its Stage IV dark energy goals, which underlie the DOE support for LSST and DESC.  Furthermore, the DDFs are crucial for the control of systematics outlined in the LSST DESC Science Requirements Document \cite{Mandelbaum2018}.  
Here we describe the constraints on the DDF parameters and their scientific motivations.  An overview of the DESC's cosmological probes is given in the DESC companion paper on the Wide Fast Deep survey (Lochner et al.~2018, hereafter DESC-WFD).  \\
{\bf Type Ia Supernovae observations guide nightly depth and cadence constraints}:
The cosmology probe that most directly benefits from the DDF mini-survey is use of SNIa to measure distances from $z=0.1$ to $z=1.0$.   As discussed in DESC-WFD, the WFD and DDF surveys enable SNIa measurement in complementary ways.  {\green The high-quality light curves from the DDFs will provide the light curve templates to fit the WFD data and extend the redshift range of the full sample. The DDF fields will provide the best host-galaxy data for SNIa (deepest LSST + ancillary spectroscopy and multiwavelength coverage), so leveraging that with the best LSST SNIa light curve data (higher signal-to-noise ratio, or SNR, and higher cadence than WFD) is imperative}.  {\it The goal for the DDF SNIa survey is to extend the redshift range of the SN Hubble diagram by a factor of $2$ past the WFD, while also improving the light-curve cadence by $2\times$.}  

\autoref{fig:SNRvsPhase} shows the nightly visit SNR, which is important for nightly detection, versus epoch for a SNIa at the `completeness edge' for $z=0.6$ for different LSST visit lengths.  The completeness edge is essentially the faintest SN at a given redshift within the normal distribution of SN parameters \cite{SK16}. {\green Redshift $z=0.6$ is chosen here because it is twice the redshift limit for WFD ($z=0.3$, as found in our study for DESC-WFD)}.  We require SNR $>10$ in 3 filters so that the light curves have sufficient quality for training the light-curve fitter and inferring cosmology \cite{Betoule14,Scolnic18b}.  This constraint defines the needed nightly depths.  Our requirement for cadence of the DDF is to be twice as frequent as that of the WFD, which is 5--8 days per filter, depending on the survey strategy (DESC-WFD).

The impact of the DDF SN survey on the Dark Energy Task Force Figure of Merit (FoM, \cite{albrecht2006}) is shown in \autoref{fig:DEFoM}. {\green The WFD+DDF FoM is $1.36\times$ that of the WFD alone}.  {\it In fact, without the DDF component, the entire LSST~DESC program could not provide the necessary FoM to be classified as a Stage IV dark energy mission (see \cite{Mandelbaum2018}).}

Another {\green benefit} of the DDF mini-survey for SNIa is the ability to stack observations of highly multiplexed spectrographs like 4MOST in the south and Subaru PFS or DESI in the north to obtain host galaxy redshifts of faint SNe (and classify live SNe), along the lines of Pan-STARRS \cite{Jones18} and the Dark Energy Survey SN program \cite{Flaugher2015,Kessler2015}} with the OzDES survey and the AAT telescope.

{ {\green \bf Season length should be $>6$ months for observing high-$z$ SNIa}}:
Because of the 2+ month timescale of an SNIa at $z=0$, and the time dilation of up to $2\times$ for redshifts up to $z=1$, season length is critical.  Changing from 4- to 6-month seasons more than doubles the number of well-measured SNIa distances.  Hence, we try to maintain a $>6$-month window for observing. {\green This is further explored in the metric section.}

{  \bf Usage of the four previously chosen deep fields and {\green selection of an additional deep field for synergy with Euclid and WFIRST}}: {\green The \texttt{baseline2018a} strategy observes 5 fields, but the final location for the fifth field has not been determined.  Four deep fields with a vast amount of ancillary data have already been chosen by LSST, so we use all of them as part of our DDF mini-survey strategy.  We recommend that the fifth deep field be chosen to have synergy} with Euclid \cite{Laureijs11} and WFIRST \cite{Spergel15} in their continuous viewing zone (which is particularly important for transient studies because of longer seasons).   The most immediate impact of overlap between the LSST DDFs and WFIRST and Euclid Deep Fields will be improved photo-$z$ {\green scatter} at the $20\%$ (Graham et~al.~2018 in prep.) to $50\%$ level \cite{Rhodes17} from $0.3<z<3.0$. The improved DDF photo-$z$ estimates will be critical for training photo-$z$ estimates in the WFD survey, where the wide Euclid NIR coverage region is too shallow to significantly improve LSST photo-$z$ on its own.  Also, the deep and high-resolution Euclid and WFIRST images will be very important for deblending, which may be  the largest WL systematic uncertainty \cite{Melchior18}. Finally, SNe observed nearly contemporaneously by both LSST and Euclid or WFIRST will allow for a {\green calibrated NIR extension at the mmag-level to the optical SN light-curves,  significantly reducing systematics from host-reddening {\green \cite{WV08,Burns18}}.}

 The two drivers for having five fields (total of 50 deg$^2$) of deep surveys are to expand on the statistics of the SNIa sample, and to test weak lensing (WL) systematics.  While a single LSST image provides enough galaxies to serve as a training sample for galaxy shape measurements \cite{BernsteinArm2016}, we recommend 3-5 such fields, spatially separated to reduce cosmic variance.  An area of at least 40 deg$^2$ is also the recommended Euclid deep field size needed for more general shear estimation methods \cite{Viola2014}.

{\bf  Co-added depth constraints after 1 and 10 years for photo-$z$ and WL systematics}:
After 1 year, we propose that the co-added DDF survey depth should exceed the 10-year WFD depth to begin validation of photo-$z$ methods.  For training samples for WL shear estimation, we propose that the total 10-year effective
exposure time (open-shutter time factoring in transparency and sky brightness) 
exceeds that in the WFD survey by at least a factor of 10 for any chosen upper bound
of seeing FWHM, at least in $r$ and $i$ bands but ideally all bands. The $10\times$ constraint is explained in \cite{BernsteinArm2016} and is critical for providing templates for characterizing the impact of blending on shear and photo-$z$ with injection simulations \cite{2016MNRAS.457..786S,2018PASJ...70S...6H}.

{\green Finally, for calibration purposes, the near-daily observations are the best opportunity for monitoring stability of the instrumental throughput.}  

{\bf \indent Allocation of 5-7\% of the full survey to DDFs}: {\green While larger allocation to the DDF mini-survey would boost the impact of this program, we need to specify the allocation range in order to limit the optimization space of this exercise and we find this range is adequate to reach dark energy cosmology goals as stated in \cite{Mandelbaum2018}.  This range is in line with the allocation used by simulations released with the white paper call.}

\begin{minipage}{\columnwidth}
\centering
\includegraphics[width=0.5\columnwidth]{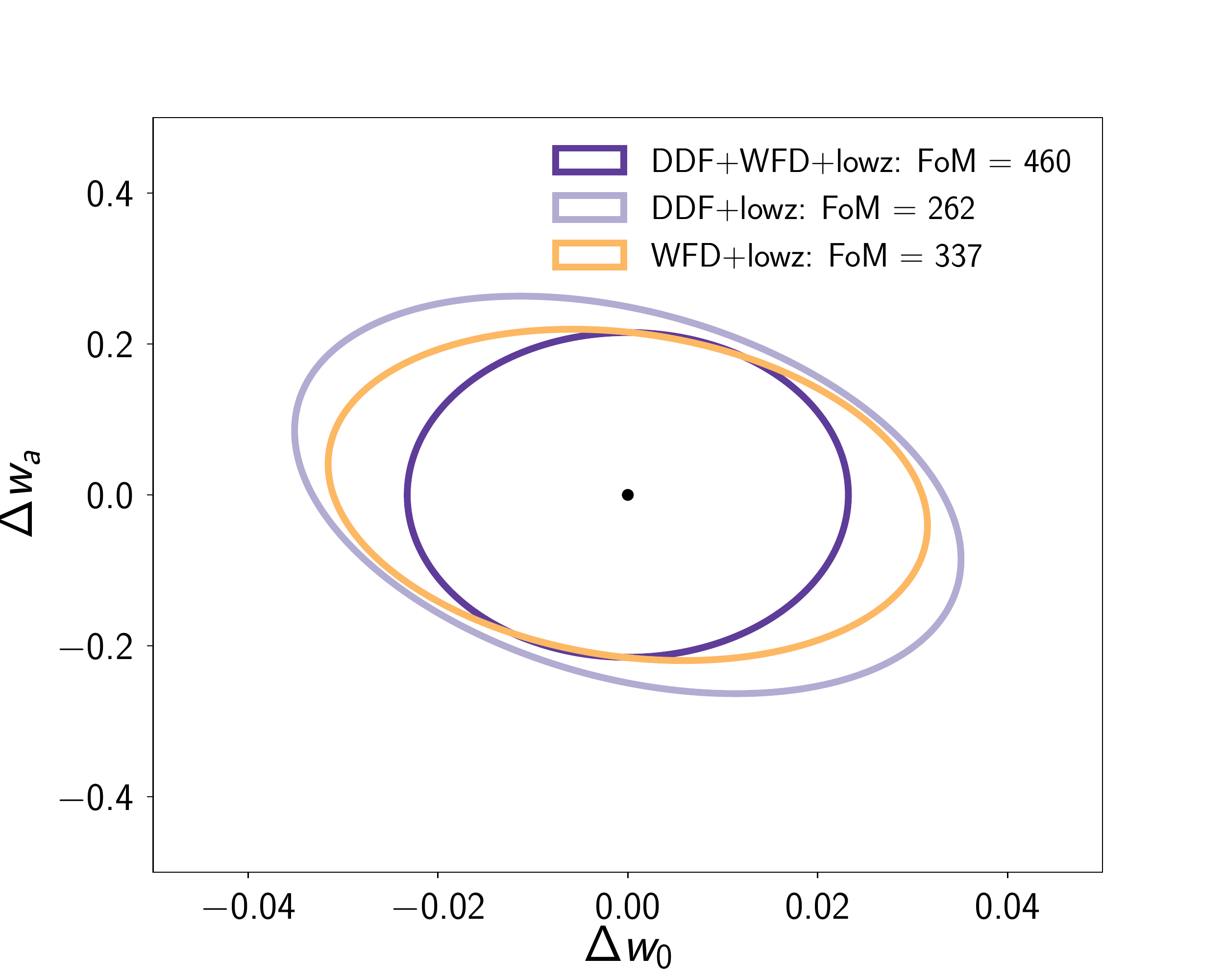}
\captionof{figure}{\green The FoM of dark energy from the LSST SNe for the 10 year DDF survey,  the WFD survey and their combination. All cases assume availability of a  low-$z$ SN sample (2400) with $z<0.1$ and a Planck-like prior \cite{Planck18}. We compare the deep fields to the WFD for the nominal cadence (\texttt{kraken\_2026}). 
The number of SNe depends on assumptions about host-galaxy follow-up, and here the DDF-only survey has $\sim1/4$ the number as WFD-only.  The mean redshifts of the DDF-only and WFD-only are $z=0.73$ and $z=0.41$ respectively. The DDF survey will enable a competitive dark energy constraint by itself, with roughly 2/3 of the constraining power of the WFD survey. Note that the FoM values are higher than those presented for SN-only in the DESC~Science Requirements Document, as we have not considered the calibrated (astrophysical) systematics in this case, to compare directly how the numbers and redshift distributions affect the FoM. }
\label{fig:DEFoM}
\end{minipage}
\begin{minipage}{\columnwidth}
\centering
\includegraphics[width=0.8\columnwidth]{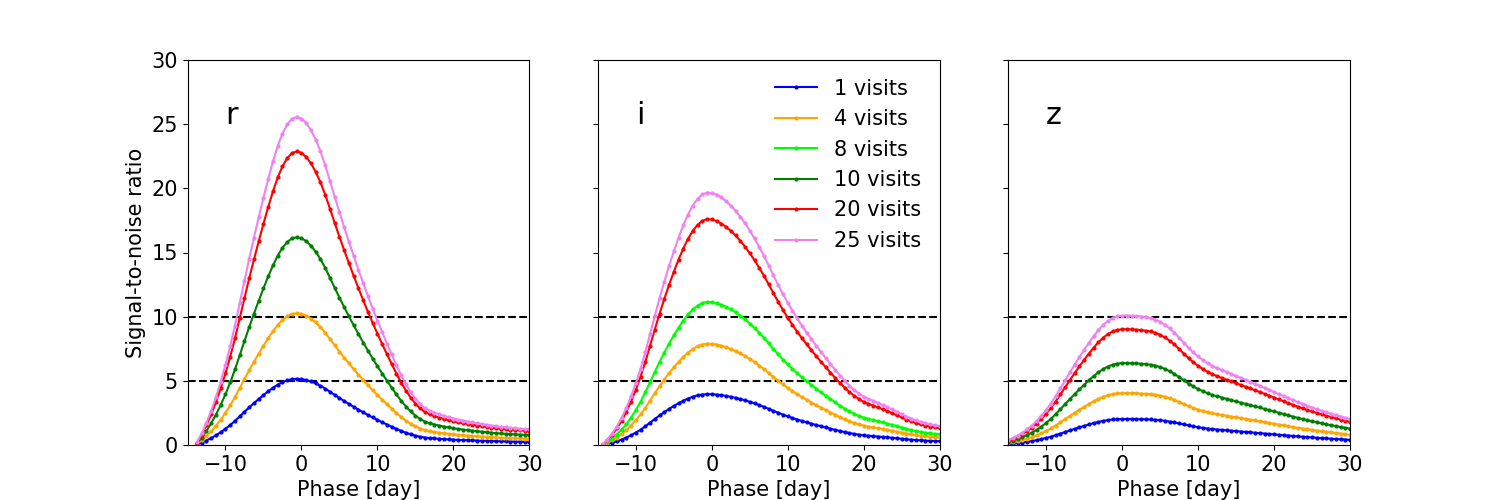}
\captionof{figure}{{\green  The SNR in $riz$ versus restframe-epoch/phase for a SNIa at $z=0.6$ with faint light-curve properties. Each curve shows the SNR for different numbers of nightly visits (of 30s each). In order to ensure SNR $>10$ in at least three bands, we require 4, 8 and 25 visits in a single night for $riz$; $y$-band SNR (not shown) is insufficient for comparable numbers of visits.}}
\label{fig:SNRvsPhase}
\end{minipage}

\vspace{.6in}
\section{Technical Description\label{sec:technical}}
\begin{footnotesize}
\end{footnotesize}
{\blue
\begin{itemize}

\item {\bf Season length to be $>6$ months:}  Here, we aim for 6.5 month seasons.  This is further explored in the metrics section. 

\item {\bf Type Ia Supernovae Observations Guide Nightly Depth and Cadence Constraints:}  Given the SNR constraints above, we would like the number of 30-s visits per night to be of 4, 8 and 25 in the $riz$ filters.  $g$ and $y$ band are less useful for this study, but are useful for photo-$z$ requirements, so we set a lower bound on number of visits to be 2 in $g$ and $4$ in $y$.  

Given the constraint on having twice as frequent visits as WFD, and a season length of 6.5 months, {\green we find an} optimum cadence is every 3 nights in each filter, and alternating between $gri$ and $zy$ every 2 of the 3 nights.  The interleaved pattern improves the light-curves for distance measurements, as discussed in the metrics section.  Furthermore, we request that if more than one night has gone by without observations of a DDF field, that field is scheduled for the next night.  This logic avoids extended gaps in cadence.

The filter allocation for $u$ is decoupled from the regular cadence of the other filters.  For $u$ band, cadence is less important for variability studies, especially for high redshift SNe. However, from photo-$z$ and WL systematic constraints above, we still require 80 visits per season per field.  We recommend $u$ and $y$ are interchanged in the filter wheel, and $u$ observations are taken during a $\sim3$ day period in dark-time.

\item {\bf Usage of the four previously chosen deep fields and {\green selection of an additional deep field for synergy with Euclid and WFIRST:}} 
To enable an additional deep field for synergy with Euclid and WFIRST, we propose Akari Deep Field South (ADFS, coordinates below).  With this addition, four of the five fields have similar RA and will be observed during roughly the same part of the year.  Because of the large quantity of ancillary data for each field, our nominal plan is to have the same observing strategy for these four fields.

However, because the COSMOS field will be observable when the others are not, we increase the cadence of observing this field, so that it is observed every day, with $gri$ one day, and $zy$ the next.  {\green This choice boosts the dynamic range of cadence and depth of our studies, and in doing so enhances the possibility of discovery of young SNe \cite{Weikang17} which may constrain SNIa systematics and also can be used to better characterize SNIa rates and our selection efficiency \cite{Kessler18}.}

This plan is visualized in \autoref{fig:cadence}.

\item {\bf Allocation of full survey to DDF to be 5-7\%:} Since this constraint is used, there is a limitation to the combination of nightly depth per filter, cadence and season length.  The constraints above fit within this total time.

\end{itemize}
\begin{minipage}{\columnwidth}
\centering
\includegraphics[scale=0.8]{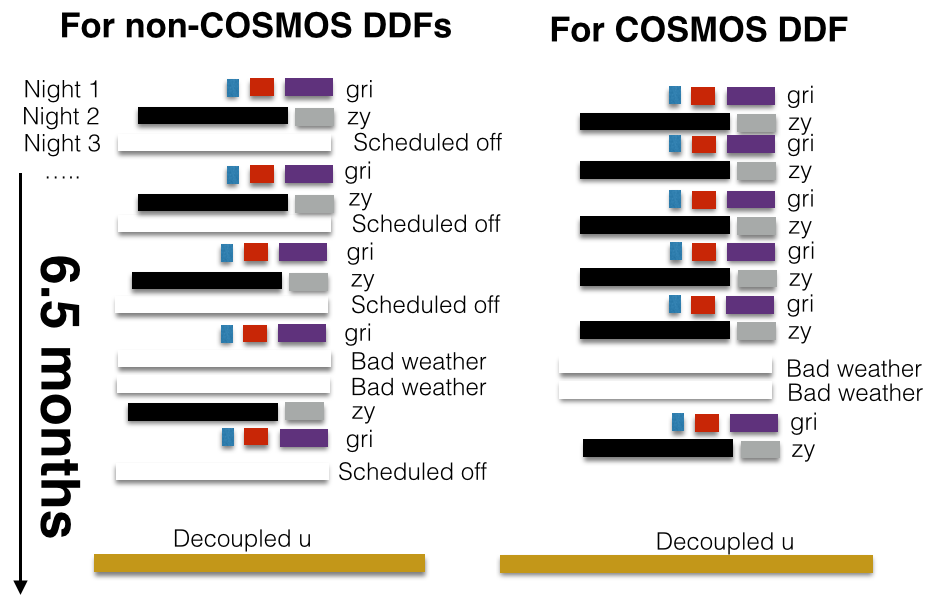}
\captionof{figure}{A visualization snapshot of our proposed cadence for a typical season for DDF (non-COSMOS on left and COSMOS on right).  The length of the bars is proportional to the number of visits per night, and the cadence (vertical slices) is in units of 1-day.}
\label{fig:cadence}
\end{minipage}

Several other white papers have elements that are synergistic to ours:
\begin{itemize}
\item {\bf Active Galaxy Science in the LSST DDFs:} {\green The footprints, cadence requirements, and total-depth requirements are similar for the AGN studies, though they would shorten the cadence to 2 days in all filters, whereas we have a similar cadence for COSMOS but 3 day for the other fields.  Additionally, our filter observations are interweaved ($gri$,$zy$) to achieve nightly depth requirements.}   
\item {\bf The Euclid and WFIRST Deep Field white papers:}  They motivate our recommended fifth field.
\item {\bf The Transients and Variable Stars Science Collaboration (TVS) GW-ToO white paper:}  The discovery of optical counterparts (kilonovae) of gravitational wave (GW) events enables a new probe to measure the Hubble constant (\cite{Holz05}, Soares-Santos et al.~2018 in prep.).
The high-level parameters for using LSST target of opportunity exposures that DESC members advocate for are similar to those being proposed by the TVS GW follow-up group. In particular, the estimates of the total time required for follow-up are similar (1 to 3 nights per year). It is important for us to follow-up all accessible GW events, even when they are, in principle, sufficiently well localized to be accessible to other instruments. This will enable collection of a uniformly selected sample, and allow us to complete our precision cosmology analysis internally within DESC.  We arrive at our estimate of time required by assuming that the LIGO/Virgo detectors will be operating at design sensitivity at the start of LSST operations (37 events per year, typically 60 deg$^2$ per event). A major upgrade is planned for 2024. This is expected to double the sensitivity range, increasing the event rate by a factor of $\sim8$. At the same time, however, the average localization area is expected to decrease by a factor of $\sim6$, so the total time required remains roughly constant.

\item {\bf The Euclid Northern Declination shallow extension:} While this is not critical for the success of DESC, it is useful for measuring dark energy, which is the goal of DESC.  Furthermore, it provides templates for Gravitational Wave ToO follow-up.
\end{itemize}

}

\subsection{High-level description\label{sec:highlevel}}
\begin{footnotesize}
{\it Describe or illustrate your ideal sequence of observations.\\}
\end{footnotesize}

{\blue
At any time in the survey, there will be as many as 4 DDFs and as few as 1 DDF (COSMOS) viewable with an airmass limit of 1.5.  When only one field is observable, we propose to increase the cadence to image $gri$ one night and $zy$ the next night for higher light-curve quality and co-added depth.

For this white paper, we define a `visit' to be `1$\times$30s' exposure in one filter and a `sequence' to {\green be} the single set of visits in all filters at one position that is repeated with a certain cadence.
The \texttt{baseline2018a} DDF program {\green observes 5 fields} and has a cadence of  265 sequences (on different nights) of 10 $g$, 20 $r$, 20 $i$, 26 $z$ and 20 $y$ visits spread over 10 years. This corresponds roughly to observations taken every 3-4 nights, if the ‘observing season’ is assumed to be 4 months long for each field.  In addition, there are 185 sequences of 40 $u$ band exposures, spread over 10 years.  In total this is 29140 visits per field, or 145700 total visits.   The total number of visits of LSST is 2.77 million, so this is 5.3\% {\green of the survey time}.  

We propose to change the cadence to be:
\begin{itemize}
    \item Night 1: $2 g, 4 r, 8 i$
    \item Night 2: $25 z, 4 y$
    \item Night 3: Nothing.
    \item Night 4: =Night 1 (repeat sequence).
\end{itemize}

We propose that the season length increases to 6.5 months, and for the COSMOS field the cadence would increase to a 2-day rather than 3-day cycle, so as to create an ultra-deep field with better time sampling.  Furthermore, we request that if one or more nights pass by without observations of a DDF, then that DDF should be observed the following night, in order to minimize gaps.  This would ensure the $2\times$ better cadence for the DDF program than WFD.

We propose for the number of $u$ visits to be 80 sequences of 10 $u$ visits.  We would like to ensure that our observations are sufficiently well-separated in time to allow for effective asteroid removal.   In \autoref{sec:requests}, we discuss observing each field every night, accumulating the same number of visits.

To calculate the total amount of survey time used, we define the total number of visits per field $=$ Filter exposure sequence in $grizy$ $\times$ Number of exposures per month $\times$ \# of months $+$ number of $u$-band sequences $\times$ number of $u$ band visits

\begin{equation}
(2+4+8+25+4) \times 30/3 \times 6.5 \times 10+80 \times 10=28750
\end{equation}

Total number of DDF visits = Number of visits per field $\times$ Number of fields (with cadence of 2 days rather than 3 for COSMOS).

\begin{equation}
28750 \times 4+28750 \times 3/2 = 158125
\end{equation}
visits, which is $5.7\%$ of the total number of LSST visits. \blue{Our attempt to reduce the number of multi-day gaps by triggering a field as soon as a night has gone by without observation will increase the relative total from our estimates (depending on weather loss), to a total of 6.1\% of survey. Furthermore,  slew and filter-change times are not included here; we estimate they could increase the total to 6.2\%.}

The cadence should be planned to be maintained over the full 10 year survey, though with a possible reassessment at some point.
}
\vspace{.3in}

\subsection{Footprint -- pointings, regions and/or constraints\label{sec:footprint}}
\begin{footnotesize}{\it 
}
\end{footnotesize}
{\blue

For the four previously-chosen deep fields, we request a center that optimizes overlap with ancillary multiwavelength data (these positions match those of the AGN DDF white paper).  Dithering is discussed below.  The positions are:

\begin{table}[h!]\centering
\begin{tabular}{|c|c|c|}
\hline
Field   & Central RA  & Central Dec   \\
Name    & (J2000)     & (J2000)       \\ 
\hline  
ELAIS-S1  & 00:37:48   & $-$44:01:30 \\
XMM-LSS   & 02:22:18   & $-$04:49:00 \\
CDF-S     & 03:31:55   & $-$28:07:00 \\
COSMOS    & 10:00:26   & $+$02:14:01 \\
\hline
ADFS     & 04:44:00    & $-$53:20:00 \\
*ADFS-A: &04:51:00 &$-$52:55:00 \\
*ADFS-B: &04:35:00 & $-$54:40:00\\
\hline

\end{tabular}
\caption{\label{tab:positions}}
\end{table}

As the XMM-LSS field is located near the bright ($V=6.5$) 
star Mira (located at $\alpha_{2000}=02$:19:20.8 and 
$\delta_{2000}=-02$:58:39.5), a study should be undertaken to evaluate moving the center of the field by roughly $0.2$ deg to avoid scattered light from Mira.  {\green There are also bright stars near CDFS and COSMOS, and judicious centering should be studied.}

Our fifth deep field is chosen to be Akari Deep Field South.  This field is planned to be a deep field for both the Euclid and WFIRST surveys, and is chosen to be in their continuous viewing zone.  To optimize Euclid synergy, we suggest potentially dividing this fifth field into two overlapping adjacent fields, to match the planned 20 deg$^2$ field of Euclid. {\blue In \autoref{sec:requests}, we propose an additional simulation that would have two DDFs for ADFS using the two potential locations given in \autoref{tab:positions}.}  As both the Euclid fields and especially the WFIRST ones are not permanently set, and contemporaneous observations are the greatest benefit of this added field, we advise that the positions of the deep fields and their cadence retain flexibility. For WFIRST synergy, there may be a need for creating new deep fields in years 7-10 of LSST to maximize contemporaneous observations.}  These five fields in \autoref{tab:positions} cover a range of declination that is accessible to spectroscopic instruments in the south (e.g., 4MOST) and north (e.g., PFS, DESI) for follow-up observations of transients and galaxies.

For dithering, we ask for rotational dithering, which will greatly reduce systematics and fill the circular 1.75 deg radius FOV.  As discussed in  DESC-WFD, the nightly rotational dithers can be performed by choosing a single RotTelPos value randomly, returning to this value after each filter change and then controlling instrument rotation such that all re-visits to individual fields are performed at the same value of RotSkyPos.  Adding small translational dithers is unnecessary for the DDFs.

\subsection{Image quality\label{sec:image-quality}}
{\blue
The image quality of the deep field observations must be at least as good as the WFD observations {\green on average}. 
To effectively serve as a training sample to mitigate WL systematics, the DDF effective exposure time must be $>10\times$ the WFD value for any chosen upper bound
of seeing FWHM, at least in $r$ and $i$ bands but ideally all bands. Finally, $95\%$ of the observations should have seeing better than
$1.2^{\prime\prime}$, \green{which is consistent with historical conditions}. 
}
\subsection{Individual image depth and/or sky brightness\label{sec:single-image-depth}}
\begin{footnotesize}{\it 
}
\end{footnotesize}
{\blue The median single-visit depths of the DDF should be similar to the ones for the WFD \texttt{baseline2018a}, which we find to be: 23.14,  24.54, 24.20, 23.65, 22.77, 21.92 in
the $ugrizy$ bands ({\blue all numbers for depths given in AB magnitudes at 5$\sigma$ SNR}).  We propose a sequence of exposures as 2 visits in $g$, 4 visits in $r$, 8 in $i$, 25 in $z$, and 4 in $y$.  The sequence depth is therefore 25.29, 
24.95, 24.78, 24.52, 
22.67 in $grizy$.}

\subsection{Co-added image depth and/or total number of visits\label{sec:co-add-image-depth}}
\begin{footnotesize}{\it  
}
\end{footnotesize}
{\blue

As discussed in \autoref{sec:motivation}, the constraint for using the DDF to study systematics in the WFD particularly for WL studies is to have $>10\times$ the number of visits in each filter after Y10.    The \texttt{baseline2018a} strategy has 62, 88, 199, 201, 180, 180 visits in $ugrizy$, respectively.  For the DDFs we propose the following distribution of visits: 800, 1300, 2600, 5200, 16250, 2600 in $ugrizy$, respectively. Therefore, our  strategy satisfies this $10\times$ constraint.  

To calculate the total depth, we scale the median depths of WFD by 2.5*log10(sqrt(visits)).    For $ugrizy$, for our 10-year program, we would have 26.8, 28.4 ,28.5 ,28.3, 28.0, 26.2.
The median \texttt{baseline2018a} WFD depth is 25.4, 27.0, 27.1, 26.4,
25.2, 24.4.  Therefore, we propose to go $\sim1.2-1.5$ mag fainter in $ugriy$ and $2.8$ mag deeper in $z$.

The COSMOS field will go 0.22 mag deeper than this in each filter.

}

\subsection{Number of visits within a night\label{sec:nvisits}}
{\blue 
The cadence is described in \autoref{sec:highlevel}. For $u$-band, we are proposing 80 sequences of 10 visits, {\green with an observing schedule dependent on when the $u$-band filter is installed in the filter wheel.}  We propose that the $y$ band is swapped with $u$ in the filter wheel at optimal times.  For the COSMOS field, we propose to increase the number of $u$-band images by a factor of $1.5$. }

\subsection{Distribution of visits over time\label{sec:visit-func-time}}
\begin{footnotesize}{\it 
}
\end{footnotesize}
{\green We stress that extending the season length to 6-6.5 months is critical due to the timescale of high-redshift SNe.  Observations with airmass up 2 are acceptable for meeting this constraint, but we find that for 6-month seasons, the DDFs can be observed with airmasses less than 1.4 and this is our recommendation. 

We would like to ensure that the DDF survey achieves the equivalent of WFD 10-year depth in the first year of operations, as this will provide photometry for a spectroscopic training set that improves the photo-$z$ estimates. The average number of visits in the WFD for \texttt{baseline2018a} after 10 years is (62, 88, 199, 201, 180, 180) in $ugrizy$. {\green Our setup would give 80, 130, 260, 520, 1625, 260 visits in 1 year for each non-COSMOS DDF and $1.5\times$ this amount for COSMOS.}

}
\subsection{Filter choice\label{sec:filter-choice}}
\begin{footnotesize}
{\it 
}
\end{footnotesize}
{\blue The order of observations in the set of $gri$ or $zy$ filters is not important.  As discussed in \autoref{sec:requests}, we are interested in simulating changing the sequence to be every two nights instead of three, and in this case reducing number of visits per night.  $u$ band is decoupled and will depend on when it is in the filter wheel - but visits should be done in large amounts as described in \autoref{sec:highlevel}.  

}  
\subsection{Exposure constraints\label{sec:exposure}}
\begin{footnotesize}
{\it 
}
\end{footnotesize}
{\blue Single visits should be consistent with the WFD survey. The DESC WFD white paper makes the case for 30s exposures instead of 2$\times$15s, so that is what we recommend here.}
\subsection{Other constraints\label{sec:other-constraints}}
{\blue It is critical that gaps between observations of the same field and the same filter stay below $7.5$ days, to follow our $2\times$ better constraint than WFD.}
\subsection{Estimated time requirement}
{\blue We propose an allocation of 6.2\% of the full LSST survey for DDF as explained above, though this is likely an overestimate due to how we account for $u,y$ bands and gap-coverage.  A more accurate total can come with a full simulation. }

\vspace{.3in}

\begin{table}[ht]
    \centering
    \begin{tabular}{l|l|l|l}
        \toprule
        Properties & Importance \hspace{.3in} \\
        \midrule
        Image quality & 1    \\
         \hspace{20pt} \emph{At least as good as WFD} & \\
       Sky brightness & 2 \\
         \hspace{20pt} \emph{Should not be prioritized over cadence} &\\
        Individual image depth & 1  \\
         \hspace{20pt} \emph{Multiple visits per night for faint SNe} & \\
        Co-added image depth & 2   \\
         \hspace{20pt} \emph{Related to \# of visits} & \\
        Number of exposures in a visit   & 2  \\
         \hspace{20pt} \emph{Same as WFD} & \\
        Number of visits (in a night)  &  1 \\ 
         \hspace{20pt} \emph{Follows filter sequence above} & \\
        Total number of visits & 1  \\
         \hspace{20pt} \emph{$>10\times$ WFD total} & \\
Time between visits (in a night) & 2 \\

           \hspace{20pt} \emph{Short enough for asteroid detection ($\sim$40 minutes)} &  \\
      
       Time between visits (between nights)  & 1  \\
        
           \hspace{20pt} \emph{Aiming for 2 day cadence} & \\
        Long-term gaps between visits & 1 \\
           \hspace{20pt} \emph{Should not be more than 6 days} & \\
        Season Length & 1 \\
           \hspace{20pt} \emph{6.5 months} & \\
         Dithering (rotational) & 1\\
        \hspace{20pt} \emph{Critical to ensure uniformity} &\\

        \bottomrule
    \end{tabular}
    \caption{{\bf Constraint rankings:} Summary of the relative importance of various survey strategy constraints (1=very important, 2=somewhat important, 3=not important). 
}
        \label{tab:obs_constraints}
\end{table}

\subsection{Technical trades\label{sec:trades}}

\begin{enumerate}
    \item 
    {\it What is the effect of a trade-off between your requested survey footprint (area) and requested co-added depth or number of visits?}\\
    {\blue \textrm ~~~~~In \autoref{sec:requests}, we suggest the possibility of having certain seasons where the cadence is significantly worse, and doing a  `rolling cadence' for the DDFs.   We do not want to decrease the proposed number of DDFs from 5; we would increase it to 6 if ADFS is broken into two.}

    \item 
    {\it If not requesting a specific timing of visits, what is the effect of a trade-off between the uniformity of observations and the frequency of observations in time? e.g. a `rolling cadence' increases the frequency of visits during a short time period at the cost of fewer visits the rest of the time, making the overall sampling less uniform.}\\
      {\blue ~~~~~We are requesting specific timing of visits. We would be open to change the number of fields we are observing at high cadence depending on the year.  For example, it is critical for ADFS that LSST observes it while Euclid or WFIRST are observing. When neither Euclid nor WFIRST is observing we could tolerate much reduced cadence.
     A better alternative to no observations on an off-year would be significantly reduced cadence (e.g., monthly) simply for calibration monitoring purposes.}
    \item 
    {\it What is the effect of a trade-off on the exposure time and number of visits (e.g. increasing the individual image
      depth but decreasing the overall number of visits)?}\\
         {\blue ~~~~~We would prefer that the exposure time is the same for DDF as for WFD to aid calibration of the WFD dataset.}
    \item 
    {\it What is the effect of a trade-off between uniformity in number of visits and co-added depth? Is there any benefit to real-time exposure time optimization to obtain nearly constant single-visit limiting depth?}\\ 
       {\blue ~~~~~It is important to our DDF strategy that the individual DDF visits are at least as good on average as the WFD visits.}
    
    \item 
    {\it Are there any other potential trade-offs to consider when attempting to balance this proposal with others which may have similar but slightly different requests?}\\ 
    {\blue ~~~~~Our DDF proposal is very similar to the AGN DDF proposal, with the main difference that we ask for more time in $z$ band.  That would be difficult to trade away because SNIa peak in $z$ band at high redshift.  A more workable trade off would be further decreasing $y$-band visits, or reducing the cadence for certain DDFs for specific years of the LSST survey.}
\end{enumerate}

\newpage
\section{Performance Evaluation\label{sec:eval}}
\begin{footnotesize}
{\it 
}
\end{footnotesize}
{\blue

\subsection{The DETF Figure of Merit\label{sec:DETF-FOM}}
{\green The DETF FoM is discussed as a metric for performance evaluation in the DESC-WFD WP.  The analog of the multi-probe DETF FoM shown in \cite{Mandelbaum2018} for \texttt{minion\_1016}, but extended to include DDF contributions to statistical and systematic uncertainties, is a work in progress.  Therefore, we rely on additional proxy figures of merit, as discussed below.}

\subsection{Number and median redshift of supernovae\label{sec:SN-z}} 

As discussed in DESC-WFD paper, a direct metric for the success of a SN
survey is the number of well-observed supernovae detected with DDF and
their redshift range. This metric is highly sensitive to changes to
observing strategy and thus is a powerful and complementary metric to
the FoM.  We have been exploring variants on our baseline strategy, multiplying cadence, filter depth, and season length by various factors to explore the consequences.  This is shown in \autoref{fig:SNzrange}.}

\begin{minipage}{\columnwidth}
\centering
\includegraphics[width=1.0\columnwidth]{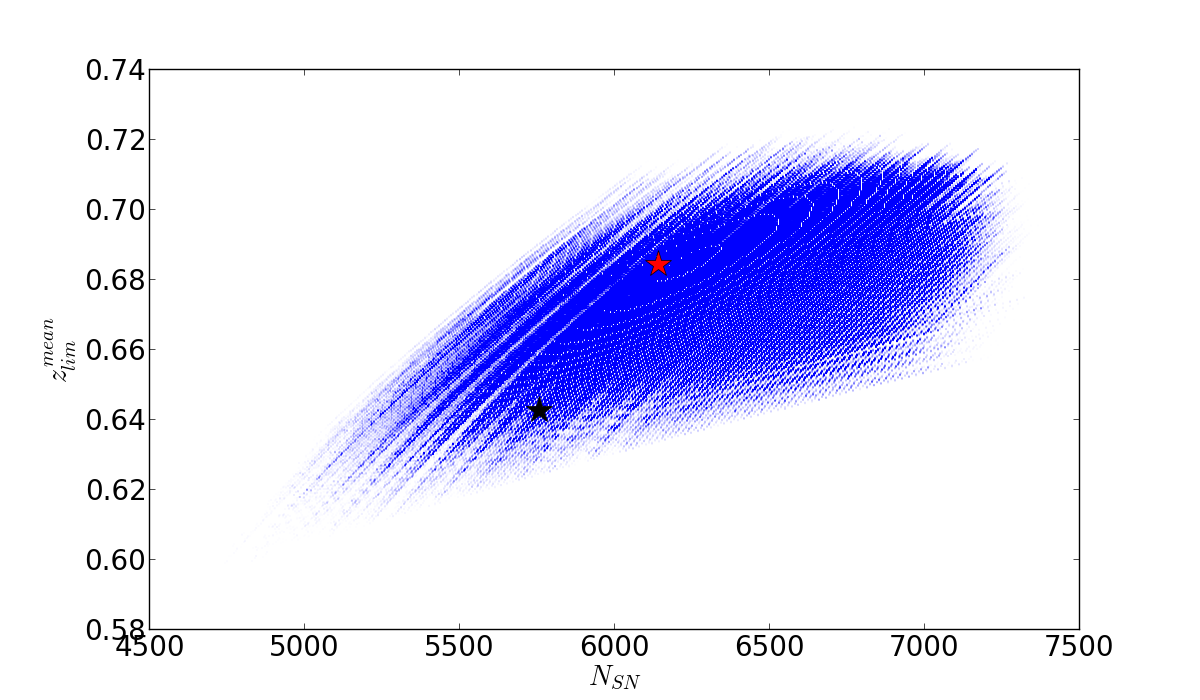}
\captionof{figure}{Using the cadence strategy requested in this white paper as our baseline, we vary visit number and season length to understand the range of scenarios for well-measured SNe (ones that pass SNR constraints given in  \autoref{sec:motivation}) versus the redshift completeness-limit for faint SNIa.  The black star shows our current strategy and the red star shows one particular variation that represents a strategy that has different cadences for different filters. All variations keep the same LSST allocation. The complexity of the program must be weighed against the optimization, which pushes scenarios towards the upper right part of the plot.  This work is ongoing. }
\label{fig:SNzrange}
\end{minipage}

\subsection{Y10 Depth\label{sec:Y10depth}} 

As discussed above, for 10 years, we propose that the total effective
exposure time (open-shutter time factoring in transparency and sky brightness) 
exceeds that in the WFD survey by at least a factor of 10 for any chosen upper bound
of seeing FWHM, at least in $r$ and $i$ bands but ideally all bands.  This metric would then pick an array of seeing values for upper bounds, and make sure the strategy passes for each one.

\subsection{Cadence regularity\label{sec:regularity}}

The quality of SN lightcurves is determined by the regularity of the
cadence delivered by the survey.  There are two simple estimators of the
cadence regularity, which are, for a given field (1) the mean of the
time interval between two (same band) consecutive inter-night visits of the field; and
(2) the frequency of 7.5 day gaps ($2\times$ better than WFD) between same-band consecutive visits --
which has to stay as low as possible.  Both metrics have been evaluated
for all cadences released so far, and will be implemented as MAF
metrics.

\subsection{Location of 5th deep field\label{eq:5thfield}}
The most simple metric to ensure the 5th DDF overlaps with the chosen Euclid and/or WFIRST deep field in their CVZ is the amount of area overlap of the LSST DDFs and Euclid/WFIRST deep fields.

\vspace{.6in}


\section{Special Data Processing\label{sec:special}}
{\blue 
For transient science, we strongly recommend a special pipeline that creates single-night co-adds (or two night co-add if sequence is staggered to be daily) in each filter for each observed DDF, performs difference imaging analysis (DIA), and creates DDF DIASource {\green (detections on individual difference images)} and DIAObject {\green  (collection of sources at a given coordinate)} catalogs. We request that these images and catalogs be updated and available through the Science Platform as soon as possible, at least before the start of the next night. We also request that Alert packets be created from the DDF DIASources and made available in a DDF Alerts Database (with appropriate flags) to facilitate ingest into Brokers (but these do not need to be streamed, just retrievable by Brokers). To facilitate DDF DIA in the first year of LSST Operations, we recommend that deep template images for each of the five DDFs be created during commissioning. Finally, we request that another special pipeline be created and run as part of Data Release processing which generates maximum depth DDF co-adds in each filter and separate Source and Object catalogs for the DDFs.

}
\section{Requests for New OpSim Runs\label{sec:requests}}
In the course of our study, we found several considerations that may impact cosmology but which require further investigation. Here we describe several new simulations we would like to request. We request that strategies with WFD, DDFs and mini-surveys be simulated together to study the resulting differences in metrics. 
\begin{itemize}
   
       \item {\bfseries Splitting ADFS into two fields:} {\green We propose to increase the area of the ADFS field by a factor of 2 but reduce the cadence by a factor of $2$ compared to the normal baseline ADFS cadence, using the two locations specified in \autoref{sec:footprint}. In this case, the number of visits per night would remain the same (which is important for detection of high-redshift SNe), but the inter-night gaps would increase by a factor of 2.  This variation would maintain the same time commitment to the DDF program.}
   
     \item {\bfseries Pushing the COSMOS field to daily cadence:} We propose to increase only the COSMOS cadence to observations of $grizy$ every day (rather than $gri$, $zy$ interweaved), keeping the same number of visits per night.  This would require $7\%$ of LSST allocation instead of $6\%$ for DDFs.
    
    \item {\bfseries Improving the cadence by a factor of $2$:} We propose to improve the cadence by a factor of $2$, {\green by halving the total exposure time per night in each filter}.

  \item {\bfseries Change snaps back to 2$\times$15s:} We would like to study the impact of this choice.  Potentially this can be used to study systematics, and we would like to know the cost/benefit of this.

    \item{\bfseries Rolling deep-field cadence:} We would like to see a simulation where not all deep fields are observed each year, and the extra time goes to increasing nightly visit amounts.
 
     \item{\bfseries Accurate seeing and weather:} In Neilsen et al. in prep., it is found that the assumed seeing has a large impact on LSST science predictions, 
    and we recommend updating the seeing and weather models (especially making some worst-case scenario weather simulations) in the next round of simulations.
 
   \item {\bfseries Variable exposure time:} Similar to the WFD, we propose to allow exposure time to vary based on observing conditions (seeing, airmass, sky brightness and transparency) to achieve more uniform depth.

\end{itemize}
\section{Acknowledgements\label{Acknowledgements}}

The DESC acknowledges ongoing support from the Institut National de Physique Nucl\'eaire et de Physique des Particules in France; the Science \& Technology Facilities Council in the United Kingdom; and the Department of Energy, the National Science Foundation, and the LSST Corporation in the United States.  DESC uses resources of the IN2P3 Computing Center (CC-IN2P3--Lyon/Villeurbanne - France) funded by the Centre National de la Recherche Scientifique; the National Energy Research Scientific Computing Center, a DOE Office of Science User Facility supported by the Office of Science of the U.S.\ Department of Energy under Contract No.\ DE-AC02-05CH11231; STFC DiRAC HPC Facilities, funded by UK BIS National E-infrastructure capital grants; and the UK particle physics grid, supported by the GridPP Collaboration.  This work was performed in part under DOE Contract DE-AC02-76SF00515.  
This document was prepared by DESC using the resources of the Fermi National Accelerator Laboratory (Fermilab), a U.S. Department of Energy, Office of Science, HEP User Facility. Fermilab is managed by Fermi Research Alliance, LLC (FRA), acting under Contract No. DE-AC02-07CH11359.

This paper has undergone internal review in the LSST Dark Energy Science Collaboration. The internal reviewers were Jonathan Blazek, Chihway Chang, Ariel Goobar, William Hartley Isobel Hook, Saurabh Jha, Nacho Sevilla Noarbe, Anže Slosar, Eli Rykoff and Peter Yoachim. 
\section{References}
\bibliographystyle{hunsrt}
\begingroup
\renewcommand{\section}[2]{}%

\endgroup

\section*{Author Affiliations}
{\parbox{\textwidth}{
\footnotesize
$^{1}$ Kavli Institute for Cosmological Physics, University of Chicago, Chicago, IL 60637, USA\\
$^{2}$ African Institute for Mathematical Sciences, 6 Melrose Road, Muizenberg, 7945, South Africa\\
$^{3}$ South African Radio Astronomy Observatory, The Park, Park Road, Pinelands, Cape Town 7405, South Africa\\
$^{4}$ Laboratoire de Physique de Clermont (LPC) , CNRS/IN2P3, F-63000 Clermont-Ferrand, France\\
$^{5}$ LPNHE, IN2P3-CNRS, 4 place Jussieu, 75005 Paris, France\\
$^{6}$ Department of Astronomy \& Astrophysics, Dunlap Institute for Astronomy \& Astrophysics, University of Toronto\\
$^{7}$ Dunlap Institute for Astronomy \& Astrophysics, University of Toroto\\
$^{8}$ Lawrence Berkeley National Laboratory, Physics Division, One Cyclotron Rd., Berkeley, CA 94720, USA\\
$^{9}$ Mullard Space Science Laboratory, Department of Space and Climate Physics, University College London, Holmbury Hill Rd, Dorking RH5 6NT, UK\\
$^{10}$ Department of Physics and Astronomy, Rutgers University, Piscataway, NJ 08854, USA\\
$^{11}$ The Oskar Klein Centre for CosmoParticle Physics, Department of Physics, Stockholm University, AlbaNova, Stockholm SE-10691\\
$^{12}$ SNSF Ambizione, Laboratory of Astrophysics, \'{E}cole Polytechnique F\'{e}d\'{e}rale de Lausanne (EPFL)\\
$^{13}$ Center for Cosmology and Astroparticle Physics, Ohio State University\\
$^{14}$ University of Chicago\\
$^{15}$ The Oskar Klein Centre for Cosmoparticle Physics, Department of Physics, Stockholm University, AlbaNova, 10691 Stockholm, Sweden\\
$^{16}$ Physics Department, Lancaster University, LA1 4YB, U.K.\\
$^{17}$ McWilliams Center for Cosmology, Department of Physics, Carnegie Mellon University, Pittsburgh, PA 15213, USA\\
$^{18}$ Kavli Institute for Particle Astrophysics and Cosmology, P.O. Box 20450, MS29, Stanford, CA 94309, USA\\
$^{19}$ Fermi National Accelerator Laboratory, Batavia, IL 60510\\
$^{20}$ NASA Jet Propulsion Laboratory, California Institite of Technology, Pasadena, CA 91109\\
$^{21}$ University of California, Berkeley \\
$^{22}$ Centro de Investigaciones Energ\'{e}ticas, Medioambientales y Tecnol\'{o}gicas (CIEMAT), Av. Complutense 40, 28040, Madrid, Spain\\
$^{23}$ Physics Department, Brookhaven National Laboratory,  Upton, NY 11973\\
$^{24}$ University of Washington\\
}
}
\end{document}